\documentclass[aps,prl,twocolumn,groupedaddress]{revtex4-1}

\usepackage{graphicx}
\usepackage{mathrsfs}

\begin{document} 

\title{Domain wall attraction and repulsion during spin-torque-induced coherent motion}
\author{E. A. Golovatski}
\affiliation{Optical Science and Technology Center and Department of Physics and Astronomy, University of Iowa, Iowa City, IA 52242}
\author{M. E. Flatt\'e}
\affiliation{Optical Science and Technology Center and Department of Physics and Astronomy, University of Iowa, Iowa City, IA 52242}

\begin{abstract}
We calculate the interaction between two magnetic domain walls during their current-induced motion.   This interaction produces a separation-dependent resistance and also a differential velocity, causing domains in motion to experience an effective attraction at large separations and an effective repulsion at short separations. In an intermediate range of currents the two domain walls will reach a natural equilibrium spacing that depends on the magnitude of the current flowing through the material.
\end{abstract}
  

\maketitle 

The effects of spin transport in inhomogeneous magnetic systems have important implications for both the understanding of fundamental physics and the development of potential applications.  Electrical generation of spin torque, which is a direct manifestation of the conservation of the angular momentum associated with spin, permits fast, localized electrical switching of magnetic domains\cite{Slonczewski1996, Berger1996, Yamanouchi2004}, electrical driving of ferromagnetic resonance\cite{Kasai2006, Sankey2006, Tulapurkar2005, Fuchs2007}, and controlled generation of coherent magnons\cite{Tserkovnyak2008, Balashov2008}.  The study of the motion of domain walls induced by this spin torque\cite{Berger1992, Dugaev2006, Gan2000, Grollier2004, Tatara2004, Thiaville2004, Thiaville2005, Yamaguchi2004, Yamanouchi2006} may lead to novel spin torque devices\cite{Allwood2005, Parkin2008, Ono2008}.  The treatment of multiple domain walls in a single system\cite{Dugaev2006.2wall,Sedlmayr2009} is a necessary next step, as many of these future devices will require the manipulation of more than one domain wall at a time.

Here we calculate spin transport properties and spin torque for a pair of $\pi$ walls  in a ferromagnetic semiconductor separated by a domain of variable size, with its magnetization oriented 180 degrees from the magnetization orientation of the far left and right leads.  Using a model Hamiltonian\cite{Vignale2002} for the coherent transport of spin-polarized carriers through a domain wall in the absence of spin-orbit interaction, and a piecewise linear approximation\cite{Golovatski2011} for the rotation of magnetization inside the domain walls, we calculate transmission and reflection coefficients, and the spin torque as a function of separation for each domain wall individually and for the system as a whole.  We find that the spin torque in each domain wall has a distinct non-trivial dependence on the separation of the walls.  The domain walls repel each other when they are very close together and attract each other when they are far apart.  This suggests that domain walls in motion may reach a natural, current-dependent, equilibrium distance and has possible implications for the motion of multiple domain walls along a ``racetrack"\cite{Parkin2008}.

\begin{figure}[h!tb]
\includegraphics[width=\columnwidth]{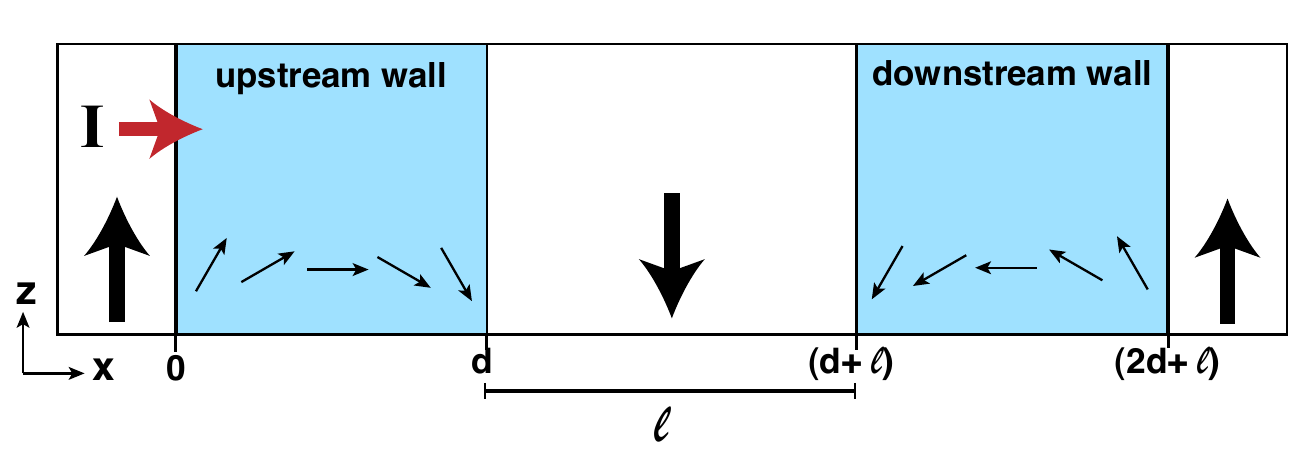}
\caption{(color online) Schematic representation of two N\'eel type $\pi$ domain walls separated by a domain with magnetization antiparallel to the magnetization in the leads.}
\label{2walls}
\end{figure}
\begin{figure}[h!tb]
\includegraphics[width=\columnwidth]{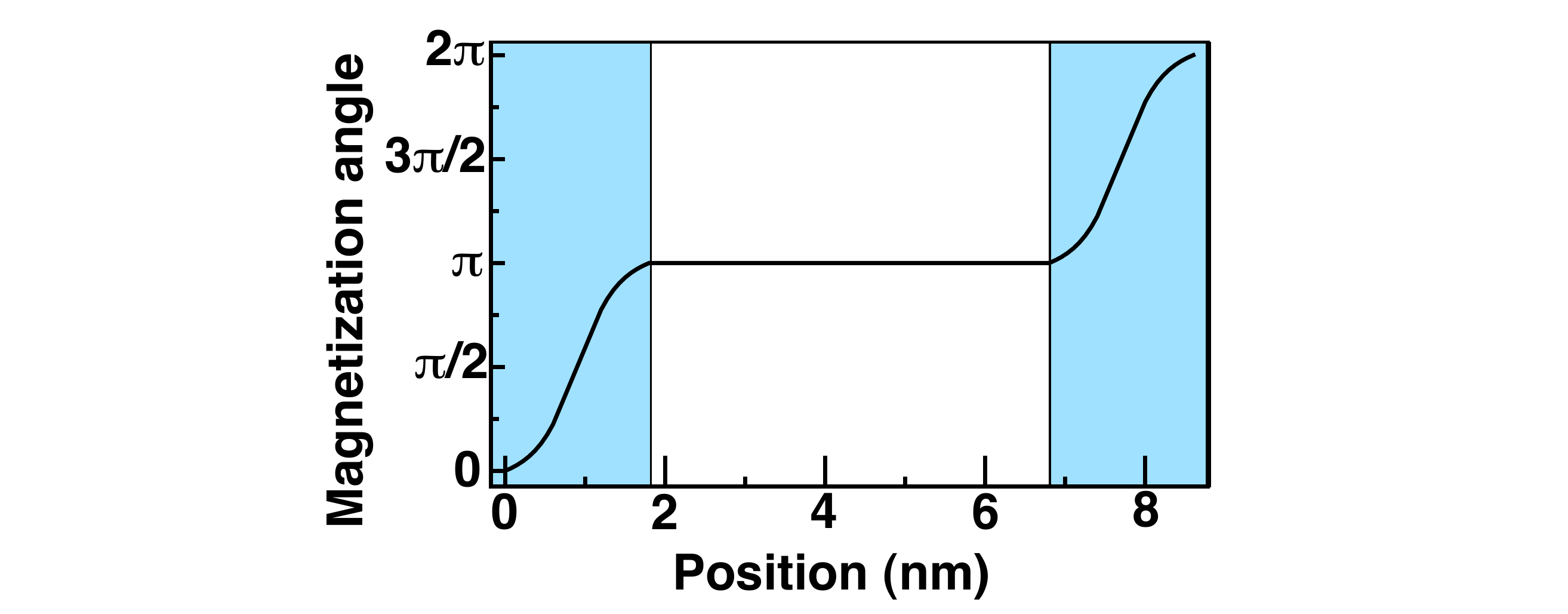}
\caption{(color online) Magnetization as a function of position for a two-domain wall system.  The domain walls here are 1.8 nm and the separation between them is 5 nm.  The blue boxes show the position of the domain walls.}
\label{magnetization}
\end{figure}

Schematics of the two-wall system are shown in Fig.~\ref{2walls}.  Two N\'eel type domain walls are separated by a domain, the magnetization of which is antiparallel to the left and right leads.    The domain walls are oriented such that they represent a full 360 degree rotation in magnetization, rather than 180 degrees and back.  Spin polarized carriers are injected from the left lead.  The exchange field for such a system can be approximated as:
\begin{equation}
{\bf B} = B_0 [\sin\theta(x) {\bf \hat x} + \cos\theta(x) {\bf \hat z}],
\label{effectivefield}
\end{equation}
where the form of $\theta(x)$ is shown in Fig.~\ref{magnetization} for a pair of 1.8 nm domain walls separated by 5 nm, and is based on a realistic form for the magnetization inside a domain wall\cite{Golovatski2011,HubertBook}.

The Sch\"odinger equation for the system with this exchange field is
\begin{equation}
\left[{-\hbar^2 \over 2m^*}{\partial^2 \over \partial x^2} - 
{\Delta \over 2}
{\left(\begin{array}{c c} \cos\theta(x) & \sin\theta(x) \\ \sin\theta(x) & -\cos\theta(x) \end{array}\right)}\right]{\left(\begin{array}{c} \psi_\uparrow \\ \psi_\downarrow \end{array}\right)} = E {\left(\begin{array}{c} \psi_\uparrow \\ \psi_\downarrow \end{array}\right),}
\label{hamiltonian}
\end{equation}
where $\Delta$ is the energy splitting between carriers of opposite spin orientation in the ferromagnetic material.

This Sch\"rodinger equation cannot be analytically solved for the $\theta(x)$ shown in Fig.~\ref{magnetization}.  We approximate the magnetization inside the domain walls as a piecewise linear function, such that $\theta_i(x) = (\phi_i (x-x_i)) / d_i$, where $\phi_i$ is the total magnetization rotation in segment $i$, $x_i$ is the leftmost position of the segment, and $d_i$ is the width of the segment.  For each linear domain wall segment, we can solve Eq.~\ref{hamiltonian} by transforming to a rotating basis\cite{Calvo1978}.  The rotation matrix:
\begin{equation}
R_i = e^{-{i \theta_i \over 2} \sigma_y} = {\left(\begin{array}{c c}\cos{\theta_i \over 2} & -\sin{\theta_i \over 2}) \\ \sin{\theta_i \over 2} &\cos{\theta_i \over 2} \end{array}\right)}
\label{rotation}
\end{equation}
defines $\psi_i = R_i\varphi_i$ and removes the $\theta$ dependence from the off-diagonal potential matrix:
\begin{equation}
R_i^{-1}{\left(\begin{array}{c c} \cos\theta_i & \sin\theta_i \\ \sin\theta_i & -\cos\theta_i \end{array}\right)}R_i = \sigma_z,
\label{rsigma}
\end{equation}
and yields a modified Schr\"odinger equation:
\begin{equation}
\left[{-\hbar^2 \over 2m^*}{\partial^2 \over \partial x^2} + {i\hbar^2\phi_i \over 2m^*d_i}{\sigma_y}{\partial \over \partial x} - {\Delta \over 2d_i^2}{\sigma_z} + {\hbar^2\phi_i^2 \over 8m^*d_i^2}\right]\varphi_i = {E \over d_i^2} \ \varphi_i.
\label{newSE}
\end{equation}
This Sch\"odinger equation can be solved analytically for the wavefunctions in each piecewise-linear segment of the domain wall.  We use transfer matrices to connect the wavefunctions in each linear segment, and then solve for the total reflection and transmission coefficients, with and without spin flip, as well as the wavefunctions for the central domain.

After obtaining the full wavefunctions for the entire system, we define a spin current density\cite{Ralph2008}:
\begin{equation}
{\bf Q} = {\hbar \over 2im^{*}}[\psi^\dagger \, {\bf S} \, (\partial_x \psi) - (\partial_x \, \psi^\dagger) \, {\bf S} \, \psi]. \label{qcurrent}
\end{equation}
The tensor {\bf Q} has a flow direction in real space as well as a direction in spin space.  As our transport model is one-dimensional, the real-space flow direction lies solely along the $\hat x$ direction, and we write {\bf Q} as a vector with components corresponding to the appropriate spin-space directions.
As this spin current is not a conserved quantity, we can then define the spin torque per unit area as the amount of spin current lost to the domain wall during transport\cite{Ralph2008}:
 \begin{equation}
{\bf N}_{DW} = {\bf Q}_{L} - {\bf Q}_{R}.
\label{ntorque}
\end{equation}
The total spin torque is then calculated by integrating the transmission and reflection coefficients over the carrier population. Calculations shown here are for a pair of 1.8 nm domain walls in a material with  a spin splitting of 100 meV and an effective carrier mass of $0.45\, m_e$ where $m_e$ is the mass of the bare electron.  The calculations are performed with a temperature of 110 K, and the carrier density is calculated to be $\sim10^{19}$ cm$^{-3}$. These parameters are representative of GaMnAs (without spin-orbit interaction), but the general qualitative features found here do not depend on the detailed quantitative parameters chosen.

\begin{figure}[h!tb]
\includegraphics[width=\columnwidth]{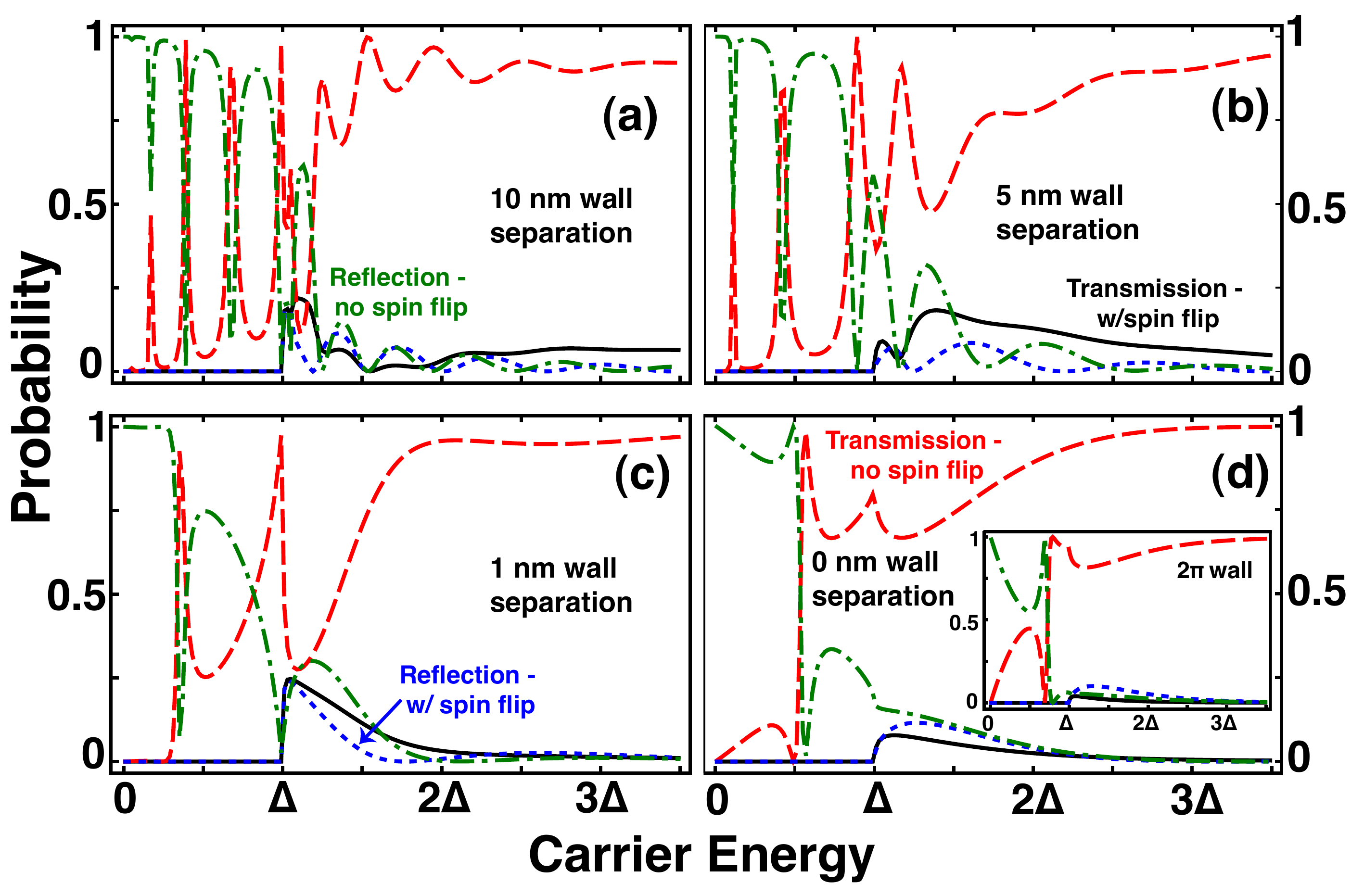}
\caption{(color online) Probabilities for transmission and reflection with and without spin flip, for a pair of 1.8 nm N\'eel $\pi$ walls with separation between them of (a) 10 nm, (b) 5 nm, (c) 1 nm, and (d) 0 nm.  Inset:  The same transmission/reflection spectrum for a 3.6 nm $2\pi$ wall. }
\label{transmission}
\end{figure}

Fig.~\ref{transmission} shows calculated probabilities for transmission and reflection of the carriers, with and without spin flip, for  these  two domain walls, for four domain wall separations.  At a distance of 10 nm (a), we see a transmission spectrum with a number of resonance peaks in the energy region below the spin-splitting $\Delta$, and largely transmission without any spin flip above $\Delta$.  This behavior is that of a spin-dependent  double barrier resonant tunneling structure.  At a distance of 5 nm (b), the resonant behavior remains, with fewer peaks.  At a distance of 1 nm (c), the peaks are even fewer, and there is additional transmission below $\Delta$ as the domain walls are getting close enough to interact more with each other.  When the two walls are brought into contact with each other (d), the transmission spectrum has many of the features of a single $2\pi$ wall (inset\cite{Golovatski2011}).

\begin{figure}[h!tb]
\includegraphics[width=\columnwidth]{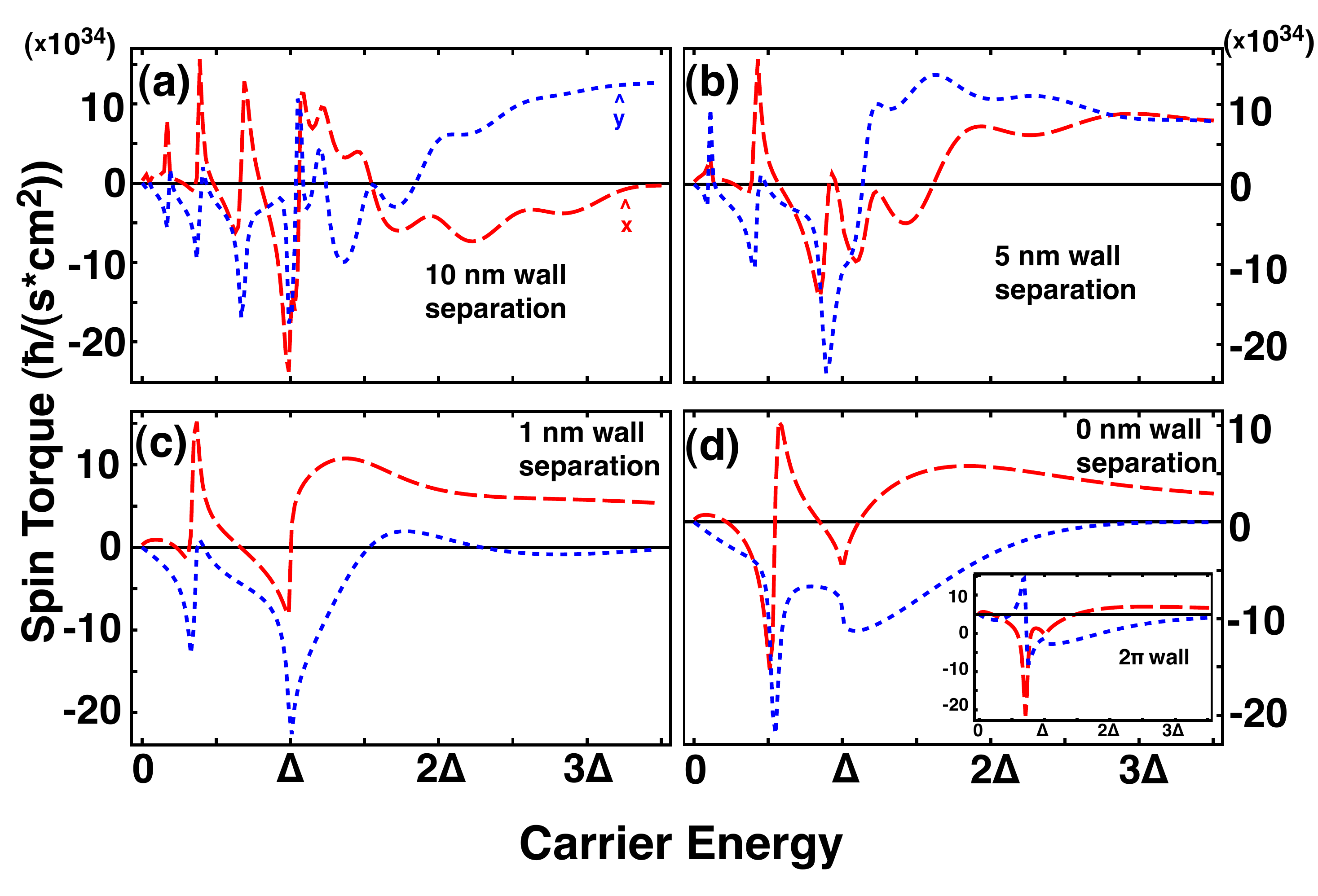}
\caption{(color online) Spin torque as a function of energy, for a pair of 1.8 nm N\'eel $\pi$ walls with separation between them of (a) 10 nm, (b) 5 nm, (c) 1 nm, and (d) 0 nm.  Inset:  The same calculation for a 3.6 nm $2\pi$ wall. }
\label{etorque}
\end{figure}

The calculated energy-resolved components (torkance) of the total spin torque acting on both walls from Eq.~(\ref{ntorque}) are shown in Fig.~\ref{etorque}.  We identify the spin torque as {\it adiabatic} (proportional to $\bf{\nabla M(r)}$, and thus parallel to $\hat x$) or {\it non-adiabatic} (proportional to $\bf{M(r)} \times \bf{\nabla M(r)}$, parallel to $\hat y$).  We see the same resonant behavior in the spin torque that we saw in the transmission spectrum for large separations of the domain walls, increasing interaction with decreasing separation, and spin torque similar to that of a $2\pi$ wall when the domain walls are brought into contact.

\begin{figure}[h!tb]
\includegraphics[width=\columnwidth]{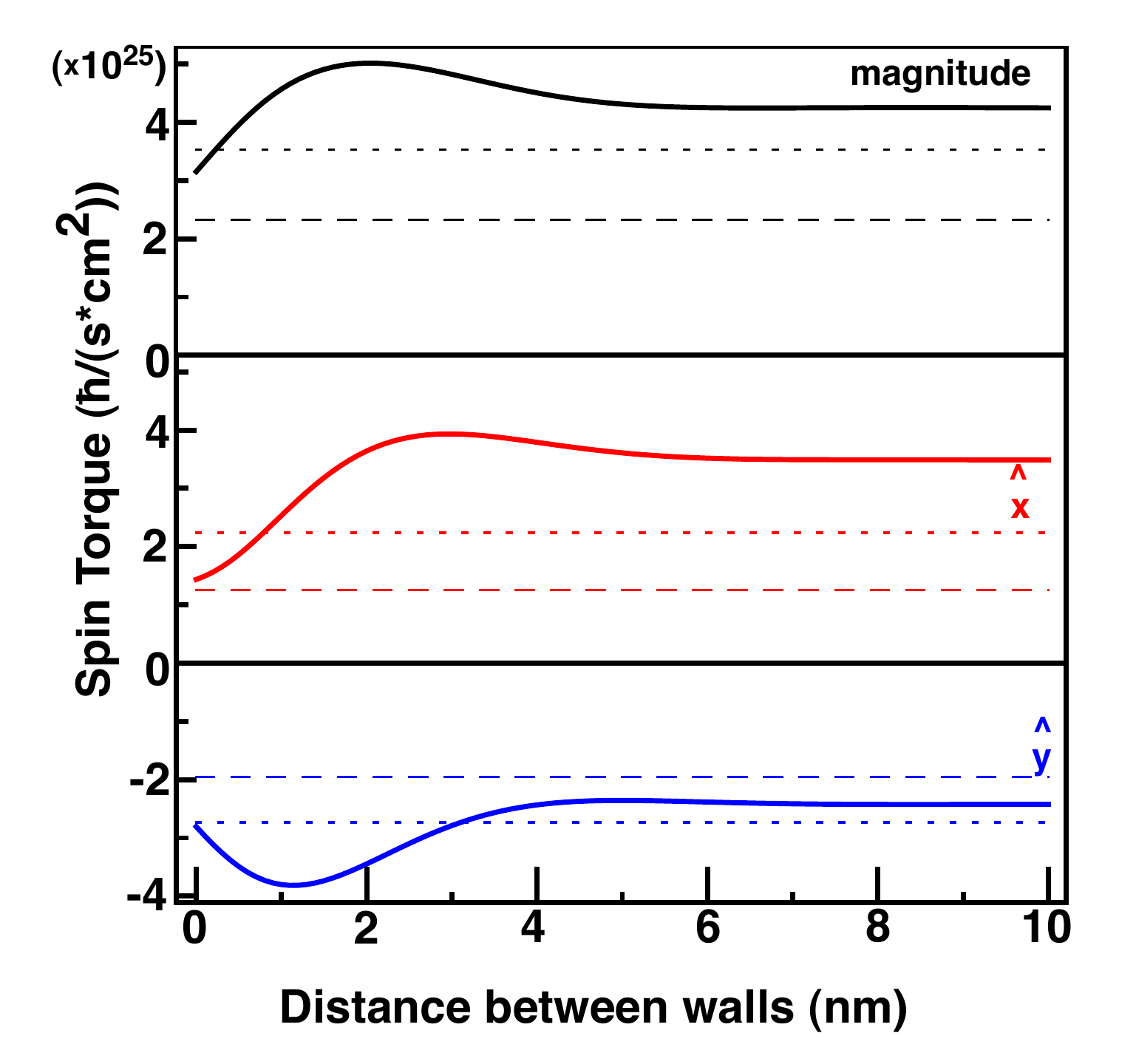}
\caption{(color online) Spin torque through the entire two wall system (solid lines).  Dashed lines are the same calculated value for a single $2\pi$ wall of 3.6 nm.  Dotted lines are twice the calculated value for a single $\pi$ wall of 1.8 nm. }
\label{dgtorque}
\end{figure}

Fig.~\ref{dgtorque} shows the total spin torque for the double wall system as a whole, as a function of the separation of the two walls, and compares it to the spin torque in individual $\pi$ and $2\pi$ walls.  The magnitude of the spin torque increases as the two walls move a small distance away from each other, and then falls to a saturation level that is higher than the spin torque in a $2\pi$ wall the same size as the two walls together, and higher than the twice the spin torque in an individual $\pi$ wall.

\begin{figure}[h!tb]
\includegraphics[width=\columnwidth]{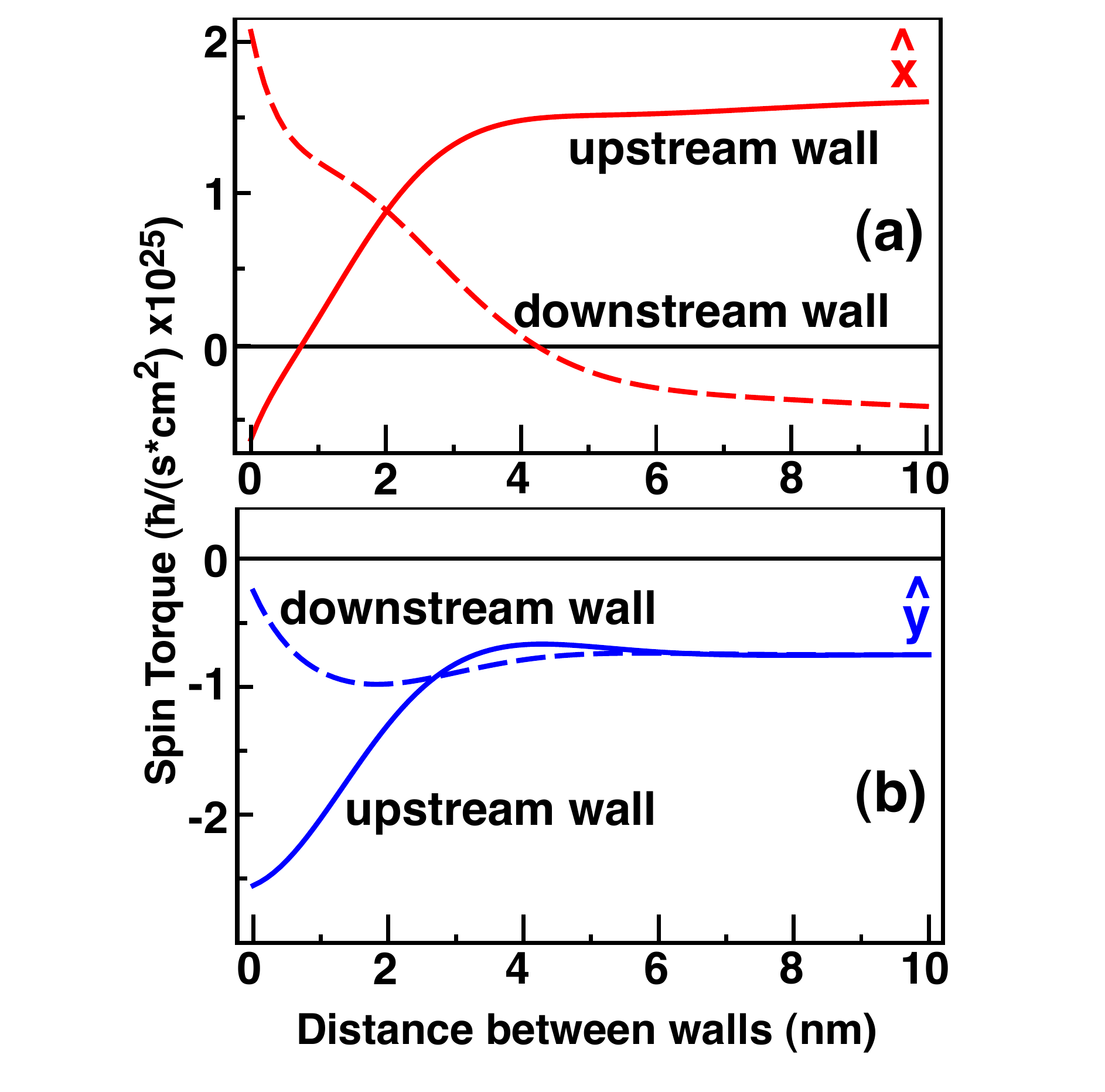}
\caption{(color online) Spin torque (a) along the direction of motion ($\hat x$) and (b) perpendicular to the direction of motion ($\hat y$) as a function of domain wall separation, for an upstream (solid line) and a downstream (dashed line) 1.8 nm N\'eel $\pi$ wall.}
\label{onetwowall}
\end{figure}

The spin current from Eq.~(\ref{qcurrent}) is a position-dependent quantity, so the torque acting on any region of space, produced by carrier motion through the region of space, can be determined from Eq.~(\ref{ntorque}). These expressions, and the values of the spin-dependent wave functions, have been used to obtain the torque acting separately on the upstream and the downstream domain walls (relative to the current direction). Fig.~\ref{onetwowall} shows the breakdown of the spin torque for each individual domain wall as a function of the separation of the two walls.  The velocity of one of these domain wall can be estimated as
\begin{equation}
v = {g \mu_b \over  M_s} N
\label{veleq}
\end{equation}
where N is the total spin torque in $\hbar$ flips per second per cm$^2$ and $M_s$ is the saturation magnetization of the material.  As this is directly proportional to the spin torque, this picture of the torque represents the motion of each domain wall.

If the motion of the domain wall can be constrained so that there isn't any motion in the $\hat y$ direction (such as is possible in certain thin films and nanowires), the motion of the domain walls is in the $\hat x$ direction, which is along the same direction as the current flow.  The spin torque in Fig.~\ref{onetwowall}(a) shows three distinct regimes of interaction between the velocities of the two walls along the $\hat x$ direction.

For very small separations between the two walls, the upstream wall moves in the $-\hat x$ direction and the downstream wall is moves in the $+ \hat x$ direction, so the separation between the walls grows. From the perspective of a coordinate system whose origin lies at the midpoint of the two walls the two walls appear to be repelling each other.  
As the separation between the walls increases, the velocity of the upstream wall slows, stops, and begins to move in the $+ \hat x$ direction.  In this second regime of interaction, the two walls move together in the $+\hat x$ direction.  As the velocity of the downstream wall decreases, this results in a separation where both walls experience equal amounts of torque, and thus move with equal velocities.  For still larger separations the downstream wall eventually slows, stops, and begins to move in the $-\hat x$ direction, while the upstream wall continues in the $+\hat x$ direction. From the perspective of a coordinate system whose origin lies at the midpoint of the two walls the two walls now appear to be attracting each other. 

The domain walls used for this calculation were 1.8 nm wide, as previous calculations showed a  high spin torque for a $\pi$ wall of this size, with a maximum around 2-3 nm width\cite{Golovatski2011}. The calculations were repeated for other widths of domain walls.  For thinner domain walls the same patterns of interaction between the two walls are found, although the equilibrium separation between upstream and downstream walls was closer together.  For much thicker domain walls (thicker than around 4-5 nm), the crossover of the velocity of each wall from positive to negative as a function of separation no longer occurs, and the interaction is repulsive for the entire range of wall separations.  The upstream wall's velocity approaches zero as the separation increases, and the downstream wall's velocity saturates to a positive value as the separation increases.

We have shown that the velocities of two separated $\pi$ domain walls depend in a sensitive and nonlinear fashion on the wall separation.  The energy-dependent electronic transmission and reflection, and the energy-resolved spin torque (torkance) change from those of a spin-dependent  double barrier resonant tunneling structure (when well separated) to those of a $2\pi$ wall (when close together).   Three interaction regimes have been found for the dependence of the differential spin torque of the two domain walls on wall separation --- repulsion, motion together, and attraction.  The existence of these three regimes indicates a stable equilibrium separation between two domain walls in the presence of current flow; thus two or more domain walls could be configured to move with a stable separation along a racetrack\cite{Parkin2008} with the application of an appropriate current.

We acknowledge support from an ARO MURI.

\end{document}